\documentclass{elsart5p}

\usepackage{graphics}
\usepackage{graphicx}
\usepackage{amssymb}
\begin{document}

\begin{frontmatter}


\title{Nonlinear paramagnetic magnetization in the mixed state of CeCoIn$_5$}
\author[AA]{H. Xiao\corauthref{Xiao}},
\ead{hxiao1@kent.edu}
\author[AA]{T. Hu},
\author[BB]{T. A. Sayles},
\author[BB]{M. B. Maple},
\author[AA]{C. C. Almasan}

\address[AA]{Department of Physics, Kent State University, Kent, OH, 44242}
\address[BB]{Department of Physics and Institute for Pure and Applied
Physical Sciences, University of California at San Diego, La
Jolla, CA, 92903}

\corauth[Xiao]{Corresponding author. Tel: (330) 672-2402 fax:
(330) 672-2959}

\begin{abstract}
Torque and magnetization measurements in magnetic fields $H$ up to
14 T were performed on CeCoIn$_5$ single crystals. The amplitude
of the paramagnetic torque  shows an $H^{2.3}$ dependence in the mixed state and an $H^{2}$ dependence in the normal state. In addition, the mixed-state magnetizations for both $H\parallel c$ and $H\parallel ab$ axes show anomalous behavior after
the subtraction of the corresponding paramagnetic contributions as linear extrapolations
of the normal-state magnetization. These experimental
results point towards a nonlinear paramagnetic magnetization in
the mixed state of CeCoIn$_5$, which is a result of the fact that
both orbital and Pauli limiting effects dominate in the mixed
state.
\end{abstract}

\begin{keyword}
torque; magnetization; paramagnetism
\PACS 71.27.+a, 74.70.Tx, 75.30.Mb, 74.25.Ha
\end{keyword}

\end{frontmatter}

The recently discovered CeCoIn$_5$ heavy fermion material is an
unconventional superconductor. A magnetic field destroys superconductivity by
coupling to either the orbits (orbital limiting) or the spins of the electrons
(Pauli limiting). The Pauli limiting effect dominates the low temperature and high field region of this system, as evidenced by
the discovered Fulde-Ferrell-Larkin-Ovchinnikov (FFLO) state \cite{Martin, Bianchi}. One
expects an unusual mixed state in which diamagnetic and
paramagnetic contributions could become anomalous since both
orbital and Pauli limiting effects are equally important.

Torque and magnetization measurements were performed on CeCoIn$_5$
single crystals in a magnetic field $H$ up to 14 T, both in the
normal and mixed states. The single crystal for which the data are presented here has a zero-field superconducting transition temperature
$T_{c0}=2.3$ K.

Angular dependent torque was measured using a piezoresistive torque magnetometer. The sample was rotated in an applied magnetic field between H $\parallel c$ axis ($\theta=0^{0}$) and H $\parallel a$ axis ($\theta=90^{0}$). Torque gives the transverse magnetic moment since
$\vec{\tau}=\vec{M}\times \vec{H}$. Typical angular dependent
torque data in the normal state are shown in Fig. 1(a). The data
can be well fitted,  as
indicated by the solid line, with
\begin{equation}
\tau_n=\tau_{n}^{max}\sin 2\theta,
\end{equation}
where $\tau_{n}^{max}$ is the amplitude of the normal-state torque. The inset to Fig. 1(a) shows that $\tau_{n}^{max} \propto H^{2}$. This magnetic field dependence of the torque is a result of the $H$ dependence of the normal-state paramagnetism of the heavy 
electrons: i.e., of $\tau_n=(1/2)(\chi_a-\chi_c)H^{2}\sin 2\theta$ \cite{Xiao}.

\begin{figure}[ht]
\begin{center}
\includegraphics[angle=0,width=0.45\textwidth]{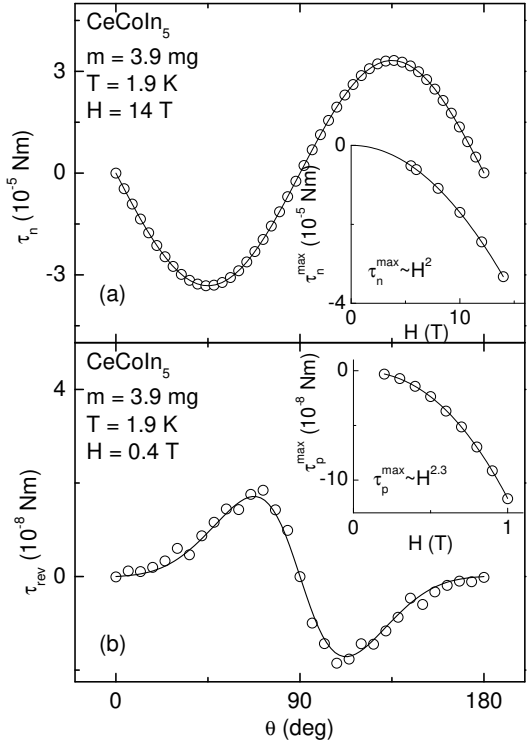}
\end{center}
\caption{Angular $\theta$ dependent (a) normal-state torque $\tau_n$  and (b) mixed-state reversible torque $\tau_{rev}$ measured at a temperature
$T=1.9$ K and magnetic field $H$ of 14 T and 0.4 T, respectively. The solid
lines are fits of the data with Eq. (1) and (2), respectively. Insets: 
$H$ dependence of the amplitude of the normal-state $\tau_n^{max}$ and paramagnetic 
$\tau_p^{max}$ torques. The solid line is a fitting curves.} \label{fig1}
\end{figure}

The heavy electrons also contribute to the mixed-state paramagnetism. The mixed-state torque displays hysteresis, so it has both reversible and irreversible parts. The reversible torque is calculated as the average of the torque measured in increasing and decreasing angle. Shown in Fig. 1(b) is the angular dependent reversible torque
measured in the mixed state, which is composed of paramagnetic and
vortex contributions; i.e., $\tau_{rev}(\theta) = \tau_{p}+\tau_{v}$, where $\tau_{v}$ is described by Kogan's model \cite{Kogan}. Hence, 
\begin{equation}\tau_{rev}(\theta) = \tau_{p}^{max}\sin2\theta +\beta
\frac{\gamma^{2}-1}{\gamma} \frac{\sin 2 \theta}
{\epsilon(\theta)} \ln\left\{\frac{\gamma \eta
H^{||c}_{c2}}{H\epsilon(\theta)}\right\},
\end{equation}
where $\tau_{p}^{max}$ represents the amplitude of the paramagnetic torque in the mixed state, $\beta \equiv \frac{\phi_0 H V}{16 \pi \mu_0 \lambda^{2}_{ab}}$ [$V$ is the volume of the sample,
$\mu_{0}$ is the vacuum permeability, $\lambda_{ab}$( =787 nm, see Ref. \cite {Xiao}) is the
penetration depth in the $ab$ plane], $\gamma = \sqrt{m_{c} /
m_{a}}$, ($m_c$ and $m_a$ are the effective mass for c and a directions, respectively), $\epsilon(\theta) = (\sin^{2}
\theta+\gamma^{2}\cos^{2}\theta)^{1/2}$, $\eta$ is a numerical
parameter of the order of unity, and $H^{||c}_{c2}$ is the upper
critical field parallel to the $c$-axis [$H^{||c}_{c2} (1.9$ K$) =
2.35$ T]. The solid line in Fig. 1(b) is the fit of the data with Eq. (2). The inset to Fig. 1(b) shows  that $\tau_{p}^{max} \propto H^{2.3}$.
The fact that the paramagnetic contribution to the mixed-state torque is not  proportional to $H^{2}$ implies
that the mixed-state paramagnetism is not a simple extrapolation of the normal-state paramagnetic magnetization, i.e. not a linear function
of $H$.

\begin{figure}
\begin{center}
\includegraphics[angle=0,width=0.45\textwidth]{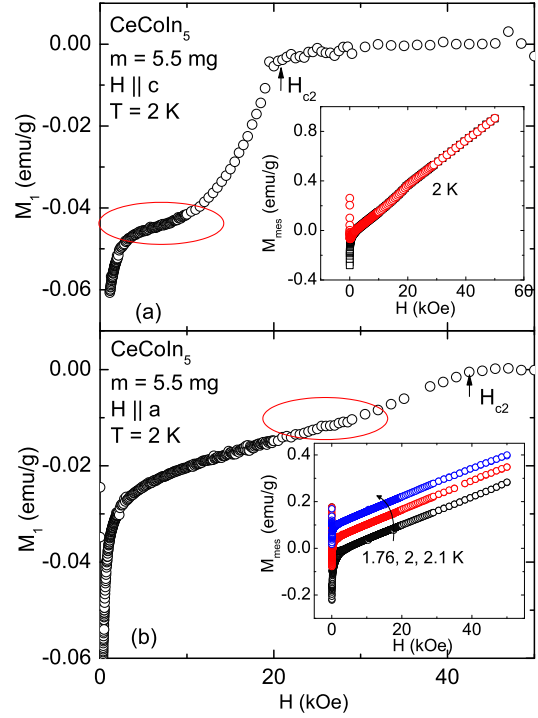}
\end{center}
\caption{Magnetic field $H$ dependence of magnetization M$_1$ for
(a) H$\parallel c$ axis and (b) $H\parallel a$ axis measured at 2 K. Insets: $H$ dependence of the measured magnetization M$_{mes}$ at $T=2$ K and $T=1.76, 2, 2.1$ K, respectively. For clarity, the curves for 2, and 2.1 K are
shifted by 0.05, and 0.1 emu/g, respectively.} \label{fig2}
\end{figure}

To get further evidence of this fact, we also performed $H$ dependent
magnetization measurements on CeCoIn$_5$.
The insets to Figs. 2(a) and 2(b) show the $H$ dependence of measured magnetization $M_{mes}$ for $H\parallel c$ and $H\parallel a$ axis, respectively.  
The diamagnetic magnetization can be obtained by subtracting the paramagnetic contribution from $M_{mes}$  in the mixed state; i.e., $M_1\equiv M_{mes}-\chi_{c,a}H$. Here we assume that the
paramagetic magnetization is just a simple extrapolation of the
normal state magnetization.  A plot of $M_1$ vs $H$ is shown in Figs. 2(a) and (b) for two $H$ directions. The obtained diamagnetic response is anomalous
for both field orientations. The $M_1(H)$ curves show kinks, as
indicated by the circles. A typical diamagnetic curve has no kinks
in it. So, magnetization measurements provide further evidence that
the assumption that the paramagnetic magnetization in the mixed state has the same
linear field dependence as in the normal state is not correct and that there must be other
contributions to the mixed-state magnetization. This conclusion is consistent
with the theoretical calculations of Adachi et al. \cite{Adachi}
for superconductors in which both Pauli and orbital limiting
effects are important.

In summary, both angular dependent torque and magnetization
measurements were performed on CeCoIn$_5$ single crystals. The
amplitude of the mixed-state torque has no longer an $H^{2}$ dependence, as
in the normal state. The $H$ dependent diamagnetic magnetization
curves are anomalous if we subtract the paramagnetic contribution
in the mixed state as an extrapolation of the normal state. Both
experiments indicate that the paramagnetism in the mixed state
is no longer a linear function of $H$. The nonlinear paramagnetic
magnetization is a result of the fact that both orbital and Pauli
limiting effects dominate in this system.

This research was supported by the National Science  Foundation
under Grant No. DMR-0705959 at KSU and the US Department of Energy
under Grant No. DE-FG02-04ER46105 at UCSD. H. X. and T. H.
acknowledge support from I2CAM through NSF grant No. DMR 0645461.\label{}


\begin{thebibliography}{99}
\bibitem{Martin} C. Martin et al., Phys. Rev. B {\bf 71} (2005) 020503{R}.
\bibitem{Bianchi} A. Bianchi et al., Phys. Rev. Lett. {\bf 91} (2003) 187004.
\bibitem{Xiao} H. Xiao et al., Phys. Rev. B {\bf 73} (2006) 184511.
\bibitem{Kogan} V. G. Kogan., Phys. Rev. B {\bf 38} (1988) 7049.
\bibitem{Adachi} H. Adachi et al., J. Phys. Soc. Jpn. {\bf 74} (2005) 2181.
\end{thebibliography}
\end{document}